\documentclass[amssymb,amsmath,jcp,twocolumn]{revtex4-2}
\usepackage{stix}
\usepackage[utf8]{inputenc}
\usepackage{amsmath}
\usepackage{hyperref}
\usepackage{graphicx}
\usepackage{comment}
\usepackage{amsfonts}
\usepackage{enumitem}
\usepackage{mhchem}
\newcommand{\beq}{\begin{eqnarray}}
\newcommand{\eeq}{\end{eqnarray}}
\usepackage{xcolor}
\usepackage{soul}
\usepackage[normalem]{ulem}

\DeclareMathOperator*{\argmin}{arg\,min}

\begin{document}
\title{Fermionic reduced density low-rank matrix completion, noise filtering, and measurement reduction in quantum simulations}
\author{Linqing Peng}
\author{Xing Zhang}
\author{Garnet Kin-Lic Chan}
\email[Corresponding author: ]{gkc1000@gmail.com}
\affiliation{Division of Chemistry and Chemical Engineering, California Institute of Technology, Pasadena CA 91125, USA}

\begin{abstract}

Fermionic reduced density matrices summarize the key observables in fermionic systems. In electronic systems, the two-particle reduced density matrix (2-RDM) is sufficient to determine the energy and most physical observables of interest. Here, we consider the possibility of using matrix completion to reconstruct the two-particle reduced density matrix to chemical accuracy from partial information. We consider the case of noiseless matrix completion, where the partial information corresponds to a subset of the 2-RDM elements, as well as noisy completion, where the partial information corresponds to both a subset of elements, as well as statistical noise in their values.  Through experiments on a set of 24 molecular systems, we find that the 2-RDM can be efficiently reconstructed from a reduced amount of information. In the case of noisy completion, this results in multiple orders of magnitude reduction in the number of measurements needed to determine the 2-RDM to chemical accuracy. These techniques can be readily applied to both classical and quantum algorithms for quantum simulations.
\end{abstract}
\maketitle

\section{Introduction}

Although quantum states live in a Hilbert space that is exponentially large in physical system size, most  information of physical interest can be captured by quantities of much reduced dimension. For time-independent fermionic observables, the relevant quantities are the fermionic reduced density matrices (RDMs)~\cite{lowdin1955quantum,davidson2012reduced}. For example, the $k$-RDM, defined as
\begin{align}
    {}^kP_{i_1i_2 \ldots i_k, i_1'i_2'i_k'} = \langle \Psi | a^\dag_{i_1} a^\dag_{i_2} \ldots a^\dag_{i_k} a_{i'_k} \ldots a_{i'_2} a_{i'_1} |\Psi \rangle
\end{align}
where $a^\dag_i, a_i$ denote fermionic creation and annihilation operators in an orbital basis, contains all information on $k$-fermion observables. We will be interested in electronic systems, where the interparticle interaction is Coulombic, and the Hamiltonian is thus of two-body form. In this case, the  2-RDM  ${}^2P_{ik, jl} = \langle a^\dag_i a^\dag_k a_l a_j\rangle$ is of particular interest, as it determines the electronic energy~\cite{coulson1960present,mazziotti2007reduced}. 

Because $\Psi$ can be quite complicated in a correlated electronic state, obtaining an accurate $k$-RDM can be expensive. Here we discuss how to obtain improved approximations to the $k$-RDM (specifically, the 2-RDM ${}^2P$, although the procedures are general) from incomplete information on its elements. We consider two types of incomplete information. The first is a noiseless setting where we have only computed a subset of the RDM elements. This situation is relevant to deterministic algorithms (or stochastic algorithms in a setting where the statistical noise is very small) when obtaining the full $k$-RDM is expensive. 
The second is a noisy setting, where the goal is to reduce the total number of measurements.
Such a noisy setting arises in both quantum Monte Carlo algorithms (as a statistical noise)~\cite{hammond1994monte,nightingale1998quantum, foulkes2001quantum,zhang201315,austin2012quantum,kent1998quantum, blunt2017density, booth2012explicitly, liu2018ab} and in quantum simulations (as a measurement shot noise)~\cite{nielsen2002quantum, kitaev1995quantum, aspuru2005simulated, cruz2020optimizing, motta2020determining, veis2014adiabatic, tilly2021reduced, braunstein1994statistical, pezze2014quantum}. In the latter case, measurement reduction~\cite{babbush2018low, wang2019accelerated, zhao2020measurement, rubin2016hybrid, huang2020predicting, huang2021efficient, zhao2021fermionic, verteletskyi2020measurement,jena2019pauli, bonet2020nearly, yen2020measuring, hadfield2022measurements, yen2023deterministic, wu2023overlapped, gokhale2019minimizing, izmaylov2019unitary,  gokhale2020n, torlai2020precise, crawford2021efficient, cotler2020quantum, google2020hartree, huggins2021efficient, shlosberg2023adaptive} is especially relevant to hybrid quantum-classical algorithms~\cite{peruzzo2014variational, yung2014transistor, shen2017quantum, mcclean2016theory, romero2018strategies, o2016scalable, nam2020ground, mccaskey2019quantum, motta2020quantum, takeshita2020increasing,hempel2018quantum, gao2021applications, parrish2019quantum, kandala2017hardware, google2020hartree, smart2021quantum, bauer2016hybrid, rubin2016hybrid, kreula2016few, yamazaki2018towards} which rely on feedback from measured quantities. The quantum shot noise will be the specific noise setting considered in this work.

Various advanced estimators have been developed to reconstruct states and processes from tomographically incomplete measurements, including the maximum-likelihood estimator~\cite{hradil1997quantum,banaszek1999maximum}, the maximum-entropy estimator~\cite{gupta2021maximal, teo2011quantum}, basis adaptive measurements~\cite{teo2020objective,kim2020universal}, and symmetry projected measurements~\cite{smart2021lowering}.
Here we will use the property that, in many applications, the RDMs are of low-rank~\cite{schwerdtfeger2012low, gidofalvi2007multireference, mazziotti2016enhanced}. Viewing the RDM as a matrix, we can then use its low-rank structure to both remove noise and/or fill in missing entries. This is a type of matrix completion or compressed sensing~\cite{cai2010singular, candes2009exact, keshavan2009matrix, keshavan2010matrix, candes2010matrix, candes2011tight, koltchinskii2011nuclear, negahban2012restricted, ge2016matrix, ji2010robust}, and in the case where all elements are available with statistical errors, a version of low-rank noise filtering \cite{richards2022remote, wall2003singular, gan2015structure, paul2000transform, kazama2001estimation,schanze2018compression, konstantinides1997noise}. 
Similar matrix completion ideas have been used in quantum state tomography to treat $n$-qubit (reduced) density matrices~\cite{gross2010quantum, gross2011recovering, flammia2012quantum, kalev2015quantum, riofrio2017experimental, cramer2010efficient, steffens2017experimentally, guctua2020fast}. Here we focus instead on the k-fermionic RDMs, and the specific matrix completion heuristics applicable to an electronic structure setting.

Matrix completion algorithms rely on a number of input parameters. We first define how such input parameters, such as the target rank, sampling method, incoherent basis, etc. can be determined in an electronic structure setting. We
further introduce simple postprocessing (or error mitigation) techniques to improve the results of the completion. 
 We then analyze noiseless and noisy matrix completion using a testbed of molecules from a subset of the G2  dataset~\cite{curtiss1991gaussian}.
In general, we find that with an optimized completion protocol, it is possible to reduce the measurement cost, either with respect to the number of elements or with respect to the number of shots, by 1-3 orders of magnitude across our dataset, while retaining a relevant accuracy to chemistry.

\section{Theory}

\subsection{Recap of matrix completion and low-rank noise filtering}
\label{sec:recapmatrixcompletion}
We  briefly recall some relevant aspects of matrix completion. For a more detailed introduction, we refer to Refs.~\cite{candes2009exact,candes2010power, candes2010matrix}. We restrict ourselves to square symmetric positive semi-definite matrices. The objective is to recover an approximation to a low-rank  $d \times d$ matrix $M$ from incomplete information about its elements. We first consider the case where we can measure the elements exactly (i.e. without noise) and the incompleteness is from measuring a subset of the elements $\Omega$. Then, given $|\Omega| \equiv N_\text{sample}$ elements of matrix $M$, 
 we can solve for a positive low-rank approximation $M^r$ through the minimization:
\begin{align} \label{eq:completion}
\begin{split}
    \min_{\tilde{M}^r} \sum_{ij \in \Omega} (\tilde{M}^r_{ij} - M_{ij})^2, \text{  s.t.  } & \text{rank}(\tilde{M}^r)=r, \tilde{M}^r \succcurlyeq 0, \\
\end{split}
\end{align}
where $\tilde{M}^r$ is the desired low-rank completion. Because we restrict to square symmetric positive definite $M$, we can use the parametrization $\tilde{M}^r=L^\dag L$ where $L$ is a real  $r \times d$ matrix, and then  perform minimization over $L$ by gradient-based techniques.

The efficiency of the above matrix completion can be discussed in terms of the fraction of sampled elements  $f_\text{sample}=N_\text{sample} / d^2$ required to obtain a given distance between $\tilde{M}^r$ and $M$,  such as the relative error (in the Frobenius norm) 
\begin{align}
    \epsilon(\tilde{M}^r, M) = \frac{||\tilde{M}^r - M||_F}{||M||}.
\end{align}
The efficiency clearly depends on the sampling scheme (i.e. the elements in $\Omega$) and how information about the matrix is distributed in its entries (the matrix coherence). Assuming a random sampling scheme, 
successful matrix completion requires information to be spread over all matrix elements. For example, a matrix with only one nonzero element can only be completed correctly if the nonzero element is sampled. 
The distribution of such nonzero information can be quantified by the coherence in terms of the singular vectors of $M$~\cite{candes2009exact, candes2010power}: for $M=U\Lambda U^T$ with $U$ a  $d \times r$ matrix,  we define the geometric coherence $\mu$ as 
\beq \label{eq:coherence}
    \mu = \frac{d}{r}\max_{1\leq i \leq d} \lVert e_i^T U \rVert^2
\eeq
where $e_i \in \mathbb{R}^n$ is the standard basis.  If all elements of $U$ have magnitude $1/\sqrt{d}$, this yields the minimum coherence $\mu=1$, while if the columns of $U$ align with the standard basis, we obtain a maximum coherence $\mu=d/r$. The number of elements required to complete the matrix successfully can be shown to increase linearly with the coherence as $O(\mu r d \ \mathrm{poly}(\log d))$~\cite{candes2009exact, candes2010power}.

In our application, we require two generalizations of the above matrix completion. The first is that  $M$ is only approximately low rank, i.e. there are $r$ singular values above some threshold, but also singular values below this threshold. Given some assumed rank $r$ in the matrix completion, we can expect the best recoverable matrix to be $M^r = U \Lambda^r U^T$ (where $\Lambda^r$ contains the $r$ largest singular values) and there is a remaining rank truncation error $\sum_{i > r} \Lambda_i^2$ where $\Lambda_i$ are the singular values in decreasing order. 
The best choice of rank $r$ is not known ahead of time. We thus discuss how $r$ can be estimated below using an independent approximate model of $M$. 

The second generalization is that we  consider matrix completion in the presence of noise. The statistical noise decreases as we increase the number of measurement shots $m$ like $1/\sqrt{m}$. The efficiency of matrix completion can be assessed as $f_m = m/m_0$ where $m_0$ is the number of shots required in some standard measurement scheme to achieve a given error in $M$.
Matrix completion is a useful technique in this context because statistical noise does not have a low-rank structure. Thus, if the noise is not too large, 
performing low-rank matrix completion  filters out the noise. There are thus two potential gains in noisy matrix completion: one from measuring fewer distinct elements of the matrix, and one from requiring fewer shots to reduce the noise. 

\subsection{The fermionic 2-RDM}
\label{sec:2rdm}

The fermionic 2-RDM, which we label $P$ for simplicity, determines the electronic energy of the system.
Given $P_{ik,jl}$, we obtain the 1-RDM
\begin{align}
D_{ij} = \sum_{kl} P_{ik,jl}\delta_{kl}/(N_\text{el}-1)
\end{align}
where $N_\text{el} = \mathrm{Tr} D$ is the number of electrons. The  electronic energy is then 
\begin{align}
E[P] = \sum_{ij} t_{ij} D_{ij} + \frac{1}{2} \sum_{ijkl} V_{ikjl} P_{ik,jl}
\end{align}
where $t_{ij}$ and $V_{ikjl}$ are the one- and two-electron integrals, respectively. We denote the size of the orbital basis by $n$. We will refer to the two-electron part of the energy as $E_2[P]$. 

The 2-RDM has a number of symmetries. In a real-orbital basis, the 2-RDM has 8-fold symmetry
\begin{align}
    P_{ik,jl} = P_{jl,ik} = -P_{ki, jl} = -P_{ik,lj} \nonumber \\
    = -P_{lj,ik} = -P_{jl,ki} = P_{ki,lj} = P_{lj,ki}
\end{align}
Further, if $S_z$ is a good quantum number, and $i\sigma, j\sigma, k\sigma, l\sigma$ label  $\alpha, \beta$ spin orbitals, $P$ has only 3 unique non-zero spin sectors: $P_{ik,jl}^{\alpha\alpha\alpha\alpha}$, $P_{ik,jl}^{\beta\beta\beta\beta}$, and $P_{ik,jl}^{\alpha\beta\alpha\beta}$. If $S_z=0$, then we further have $P_{ik,jl}^{\alpha\alpha\alpha\alpha} = P_{ik,jl}^{\beta\beta\beta\beta}$. $P_{ik,jl}^{\alpha\alpha\alpha\alpha}$, $P_{ik,jl}^{\beta\beta\beta\beta}$ have 8-fold symmetry, thus it is sufficient to consider symmetric matrices $P_{i>k, j>l}^{\sigma\sigma\sigma\sigma}$ of dimension $d \times d$ where $d = n(n+1)/2$ and $n$ is the number of spatial orbitals.   $P_{ik,jl}^{\alpha\beta\alpha\beta}$ has only 2-fold symmetry $P_{ik,jl} = P_{jl,ik}$ and is thus represented by a  $d\times d$ symmetric matrix with $d=n^2$. 
We will only sample or measure unique elements (e.g. only the lower triangular part of $P$), and non-zero spin sectors, in our completion tests below, although for simplicity, we will refer to all spin sectors collectively as $P$.

\begin{figure}[htbp]
    \centering
    \includegraphics[width=\columnwidth]{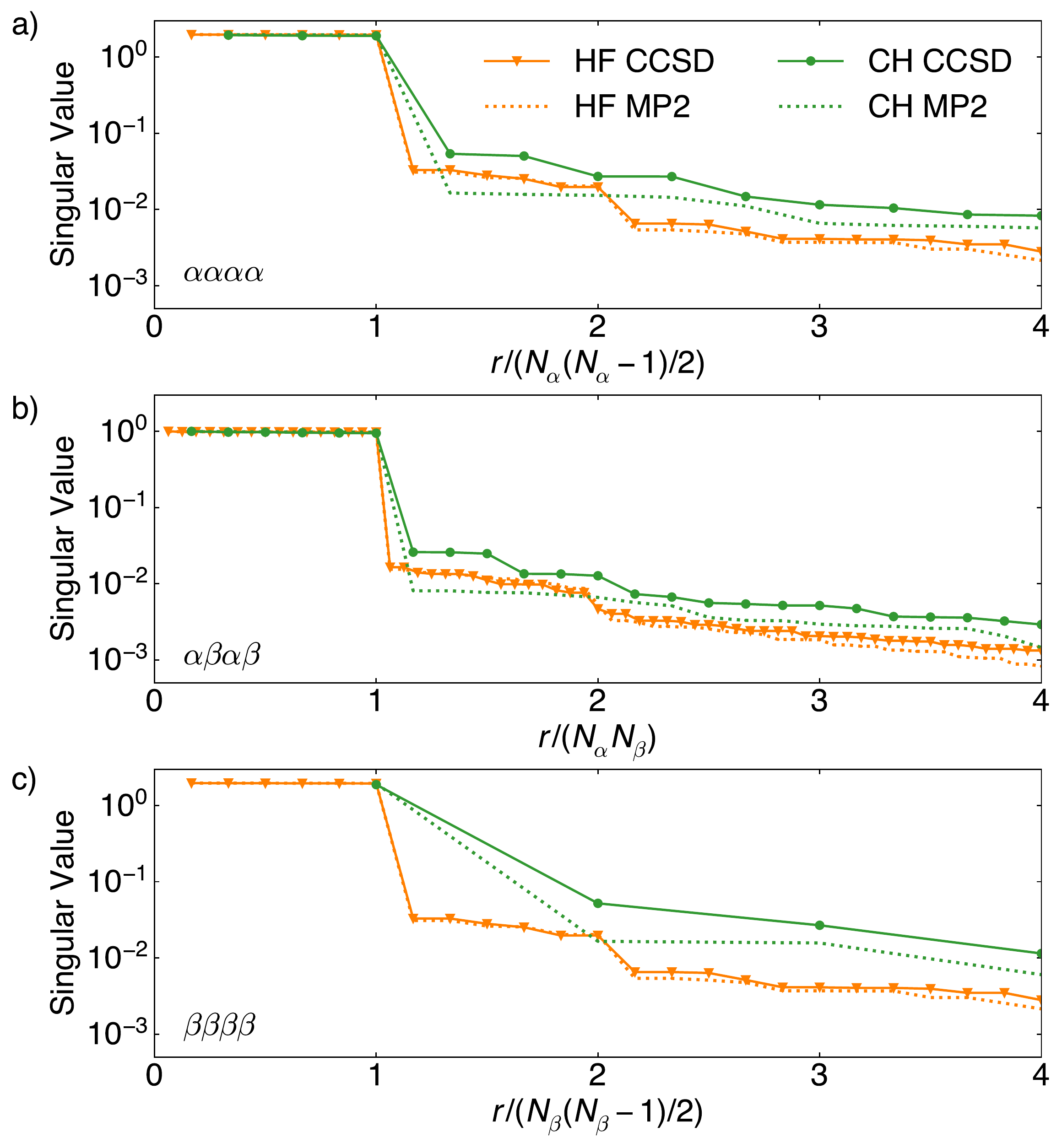}
    \caption{Singular values of the a) $P^{\alpha\alpha\alpha\alpha}$, b) $P^{\beta\beta\beta\beta}$, and c) $P^{\alpha\beta\alpha\beta}$ sectors of the (unrestricted) CCSD (solid) and MP2 (dash) 2-RDMs of the HF and CH molecules in the cc-pVDZ basis. The $x$-axis is the rank $r$ divided by the rank of the corresponding Hartree-Fock 2-RDM. }
    \label{figure:rank}
\end{figure}

The maximum rank $r$ is $d$. For orientation, if one assumes the Hartree-Fock density matrix, where 
\begin{align}
P_{ik,jl} = D_{ij}D_{kl} - D_{il}D_{kj}
\end{align}
and $D$ is idempotent, then $r(P_{ik,jl}^{\sigma\sigma\sigma\sigma}) =N_{\sigma} (N_{\sigma}-1)/2$ and  $r(P_{ik,jl}^{\alpha\beta\alpha\beta})=N_{\alpha} N_{\beta}$ where $N_\alpha$ and $N_\beta$ are the number of spin-up and spin-down electrons, respectively. These are the minimum ranks for an electronic system: if there are electron correlations, the rank of the 2-RDM increases. In Fig.~\ref{figure:rank}, we show the singular values of the spin-components of $P$ for two models of electron correlation: coupled cluster singles and doubles (CCSD) and second-order M{\o}ller Plesset perturbation theory (MP2)~\cite{purvis1982full,moller1934note, helgaker2013molecular}. In both models, the singular value spectrum contains large singular values, corresponding to the Hartree-Fock piece of the 2-RDM. 
Beyond these, the singular values decay approximately exponentially~\cite{giesbertz2013natural}.

\subsection{Noiseless completion of the 2-RDM}
\label{sec:noiselesscompletion}

We first consider the noiseless completion of the 2-RDM, where we have an incomplete sampling of the elements.
To define the minimization problem in Eq.~\ref{eq:completion} concretely, as described in Sec.~\ref{sec:recapmatrixcompletion}, we must specify (i) how to sample the elements,(ii) the estimated matrix rank $r$, and (iii) the number of elements to sample.

While there are procedures to estimate the approximation rank $r$ on the fly~\cite{keshavan2009gradient, wen2012solving}, here we use a simpler process that is likely available in many applications. Recall that we wish to use matrix completion in a setting where obtaining the elements of $P$ is expensive. We can determine a less accurate  model 2-RDM $P_M$ via a cheaper procedure, and use the model to determine the optimal sampling, choice of rank $r$, and the number of elements to sample. We define the rank $r$ model approximation  ${P^r_M} = U \Lambda^r U^\dag$ and choose $r$ such that
\begin{align}
\epsilon(P_M^r, P_M) < \kappa \epsilon_0 
\label{eq:rankthreshold}
\end{align}
where $\epsilon_0$ is our target completion error, and $\kappa$ is an empirical constant to account for the fact that our final error includes not only the rank truncation error arising from Eq.~\ref{eq:rankthreshold} but also a completion error from incomplete sampling. 
Here we use $\kappa = 1/2$. 

\begin{figure}[htbp]
    \centering
    \includegraphics[width=\columnwidth]{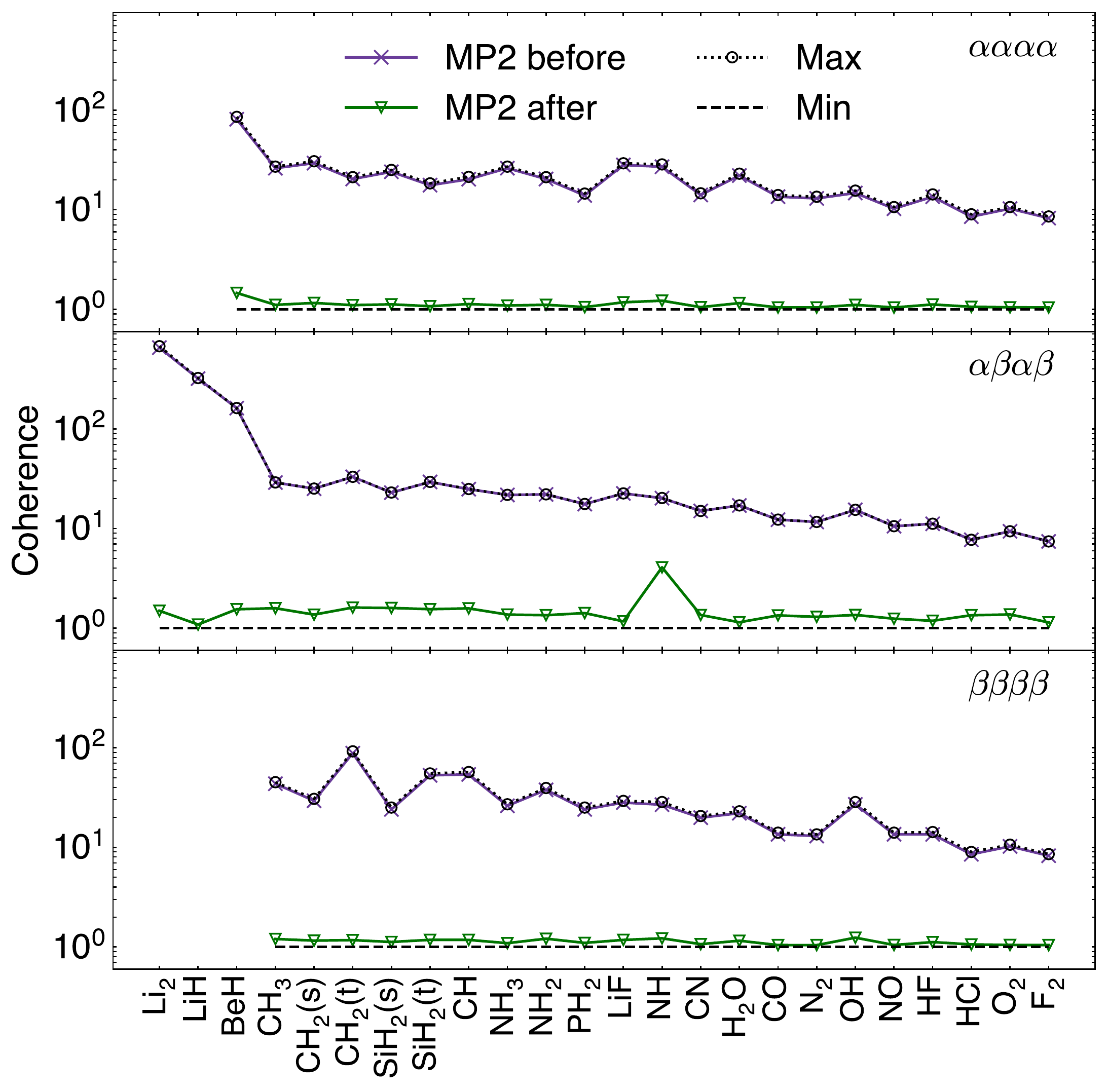}
    \caption{Coherence of the MP2 model 2-RDM in the cc-pVDZ basis. In the canonical MO basis, the coherence (purple) is close to the maximal coherence $\mu=d/r$} (black circle). 
    After random orbital rotations, the coherence diminishes significantly, approaching the minimal coherence $\mu=1$ (black dashed line). Note that after discarding the core orbitals (see main text) Li$_2$, LiH, and BeH only have a non-zero $P^{\alpha\beta\alpha\beta}$ sector. 
    \label{figure:coherence}
\end{figure}

Next, we consider the element sampling. Since we have a model available, one might consider sampling elements in the descending order of magnitude of elements of $P_M$ in a basis such as the canonical molecular orbital (MO) basis. However, for the smaller elements necessary to complete $P$ to chemical accuracy in energy, we observe a significant difference between our model $P_M$ and $P$. Thus, sampling in this order does not give a favorable completion efficiency. 
 As a result, we instead use uniform random sampling of the elements, which is efficient if the matrix is not very coherent. To minimize the coherence of the 2-RDM, we 
optimize orthogonal matrices $C$ such that the coherence of  $[P'_M]_{pqrs} = \sum_{ijkl} C_{pi}C_{qk}[P_M]_{ikjl}C_{jr}C_{ls}$ is minimized. See \ref{sec:computationaldetails} for practical implementation.
Fig.~\ref{figure:coherence} shows the reduction in the coherence of a model MP2 $P_M$ after such orbital rotations. 
 
\begin{figure}[htbp]
    \centering
    \includegraphics[width=\columnwidth]{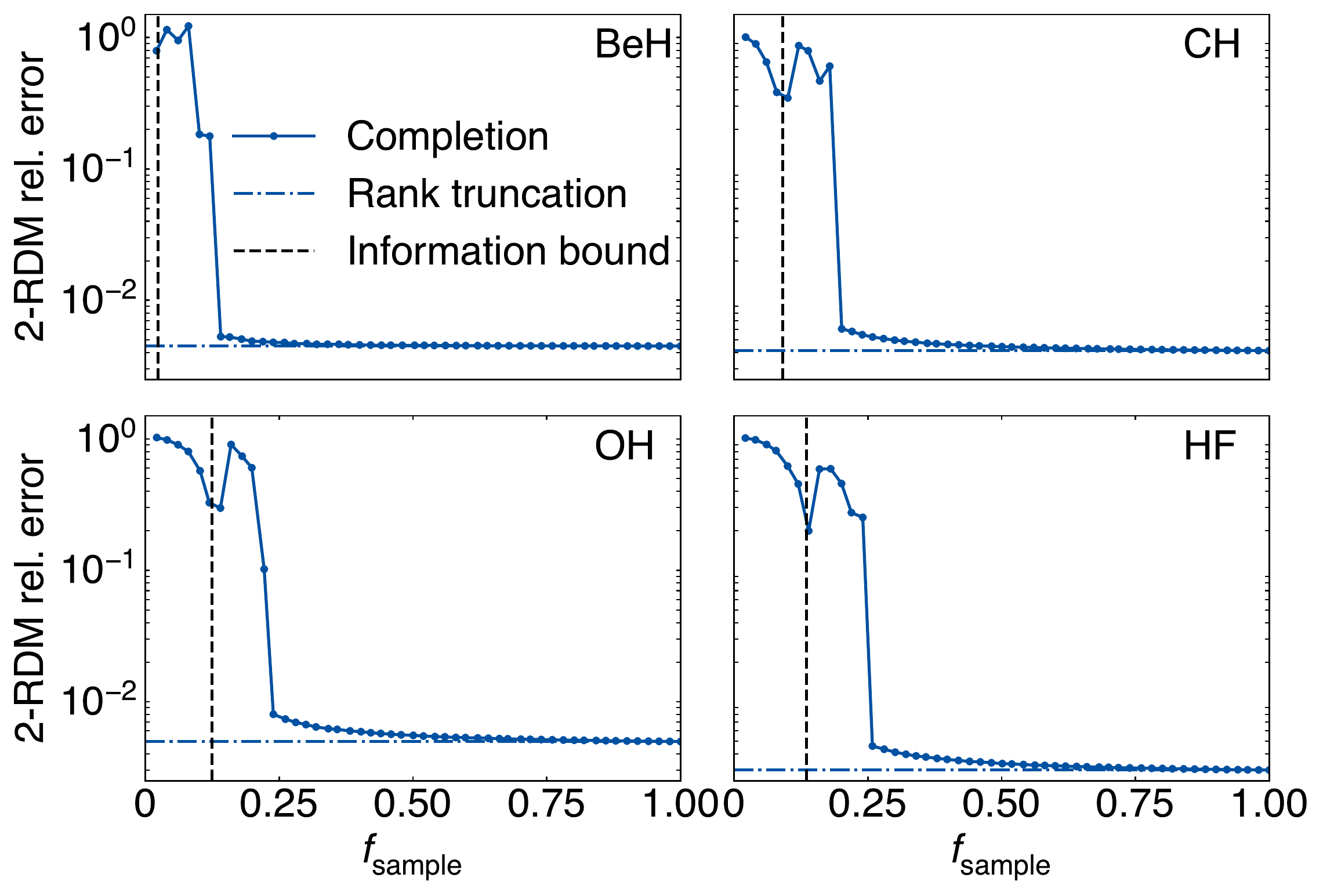}
    \caption{Completion errors of BeH, CH, OH and HF MP2 $P^{\alpha\alpha\alpha\alpha}$ RDMs in the cc-pVDZ basis as a function of the fraction of sampled elements $f_\text{sample}$. The completion rank is chosen according to Eq.~\ref{eq:rankthreshold}.  The theoretical information bound of $(2rd-r^2+r)/(d(d+1))$ is the ratio of degrees of freedom in a rank-$r$ symmetric $d\times d$ to that of a rank-$d$ symmetric matrix.}
    \label{figure:obsTrend}
\end{figure}

To estimate $f_\text{sample}$ (the fraction of elements to sample),  we perform matrix completion on the model $P_M$ for the specified rank $r$, and coherence optimized orbitals, and choose $N_\text{sample}$ so $\epsilon(\tilde{P}_M^r, P_M) < \epsilon_0$. An example of such a model matrix completion is shown in Fig.~\ref{figure:obsTrend}. As the fraction of sampled elements increases, the completion error saturates at the rank truncation error. However, there is an unusual feature where the completion error rises near the theoretical information bound (the number of elements needed to exactly complete a symmetric matrix with the exact rank of $r$). This appears related to the approximate low-rank nature of $P_M$, and the non-trivial feature which is difficult to describe purely theoretically illustrates the value in having an explicit model $P_M$ to determine the parameters of matrix completion.

\subsection{Measuring the 2-RDM in the quantum setting}

We now consider the problem of measuring the 2-RDM with noise, which we will take to arise from quantum measurements. 
We choose a Jordan-Wigner encoding of fermions and assume we are measuring Pauli operators. The expectation value of strings of Pauli operators can then be converted to fermion expectation values. For a quartet of fermion labels $i,j,k,l$, the 3 fermion expectation values not related by permutational symmetry $\langle a^\dag_i a^\dag_k a_l a_j\rangle$, $\langle a^\dag_i a^\dag_j a_l a_k\rangle$, and $\langle a^\dag_i a^\dag_l a_j a_k \rangle$
are each determined from linear combinations of the expectation values of 8 Pauli strings
\begin{align}
\label{eq:paulistring}
\langle a^\dag_i a^\dag_k a_l a_j\rangle, &\langle a^\dag_i a^\dag_j a_l a_k\rangle, \langle a^\dag_i a^\dag_l a_j a_k \rangle  \Leftrightarrow \nonumber \\
& \langle X_i \cdot X_k X_l \cdot X_j \rangle, 
\langle X_i \cdot X_k Y_l \cdot Y_j \rangle, \nonumber \\
& \langle X_i \cdot Y_k X_l \cdot Y_j \rangle, 
\;\;\langle X_i \cdot Y_k Y_l \cdot X_j \rangle, \nonumber \\
& \langle Y_i \cdot X_k X_l \cdot Y_j \rangle, 
\;\;\langle Y_i \cdot X_k Y_l \cdot X_j \rangle, \nonumber \\
& \langle Y_i \cdot Y_k X_l \cdot X_j \rangle, 
\;\;\langle Y_i \cdot Y_k Y_l \cdot Y_j \rangle, \nonumber \\
\end{align}
where $\cdot$ indicates additional possible $Z$ operators in between the $i,j,k,l$ indices. 
(Certain simplifications arise if any of the fermion indices $i,j,k,l$ are the same; we can reconstruct such fermion expectation values using Pauli strings containing $Z$ operators. We use such simplifications in our implementation and counting below.

Because the quantum state is not in a simultaneous eigenstate of all the measured operators, there will be statistical errors in (some of) the measurements. Although there exist a variety of techniques to minimize the number of measurement settings by grouping simultaneously measurable operators~\cite{bonet2020nearly,huang2020predicting, verteletskyi2020measurement}, we use the straightforward approach of independently measuring each Pauli string and leave potential improvement by grouping to future work. 
Thus we sample fermionic terms in sets of 3 in Eq.~\ref{eq:paulistring}, each set associated with a $i,j,k,l$ quartet and reconstructed from the same 8 Pauli strings.
The measurement variance for the Pauli string $Q$ is then obtained from the binomial distribution as $\sigma^2 = (1+\langle Q \rangle)(1-\langle Q \rangle)/m_\mathrm{Q}$ where $m_\mathrm{Q}$ is the number of measurements of the string.

\begin{figure}[htbp]
    \centering
    \includegraphics[width=\columnwidth]{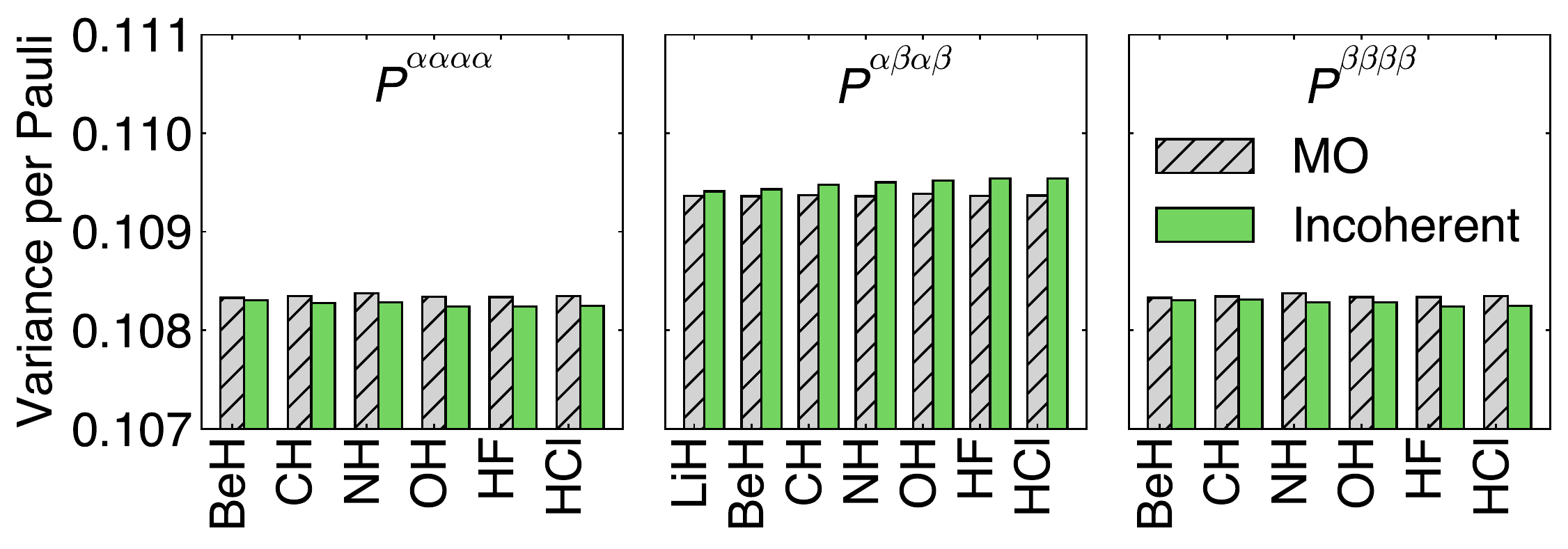}
    \caption{Average variance per Pauli string of MP2 2-RDMs in the aug-cc-pVDZ basis from a quantum measurement where each Pauli term is measured with 1 shot. The gray bars labeled ``MO'' denote measurements in the canonical MOs; the green bars labeled ``incoherent'' denote measurements in the coherence minimized orbital basis.} 
    \label{figure:variance}
\end{figure}

To define the efficiency of matrix completion, we first need to define a ``standard'' measurement procedure, where no matrix completion is performed. In this scheme, all Pauli strings required for the fermionic 2-RDM are measured 
with the same number of shots yielding a noisy $\tilde{P}$. To estimate the total number of shots $m$ required, we measure $P$ in the coherence minimized orbital basis, and choose $m$ such that $\epsilon(\tilde{P}, P) <  \epsilon_0$. 
In our tests,  
coherence minimization does not appreciably change the $m$ required, as illustrated on the model $P_M$ in Fig.~\ref{figure:variance}. 

\subsection{Noisy completion of the 2-RDM}

We now discuss matrix completion when measurements include statistical errors. Given our model $P_M$, we use the same rank estimation procedure and coherence minimization procedure as in the noiseless setting. However,
we need a different procedure to determine the number of (sets of) fermionic elements $N_\text{sample}$ to measure, as the actual cost we wish to optimize is related to the total number of measurement shots $m$.  
For simplicity, we assume each Pauli string is being measured with the same number of shots.
For a given $m$ we should then search over $N_\text{sample}$ to find the number of fermionic elements to measure that complete $P_M$ with the lowest completion error; we then increase $m$ until the completion error is below $\epsilon_0$.   

\begin{figure}[htbp]
    \centering
    \includegraphics[width=1\columnwidth]{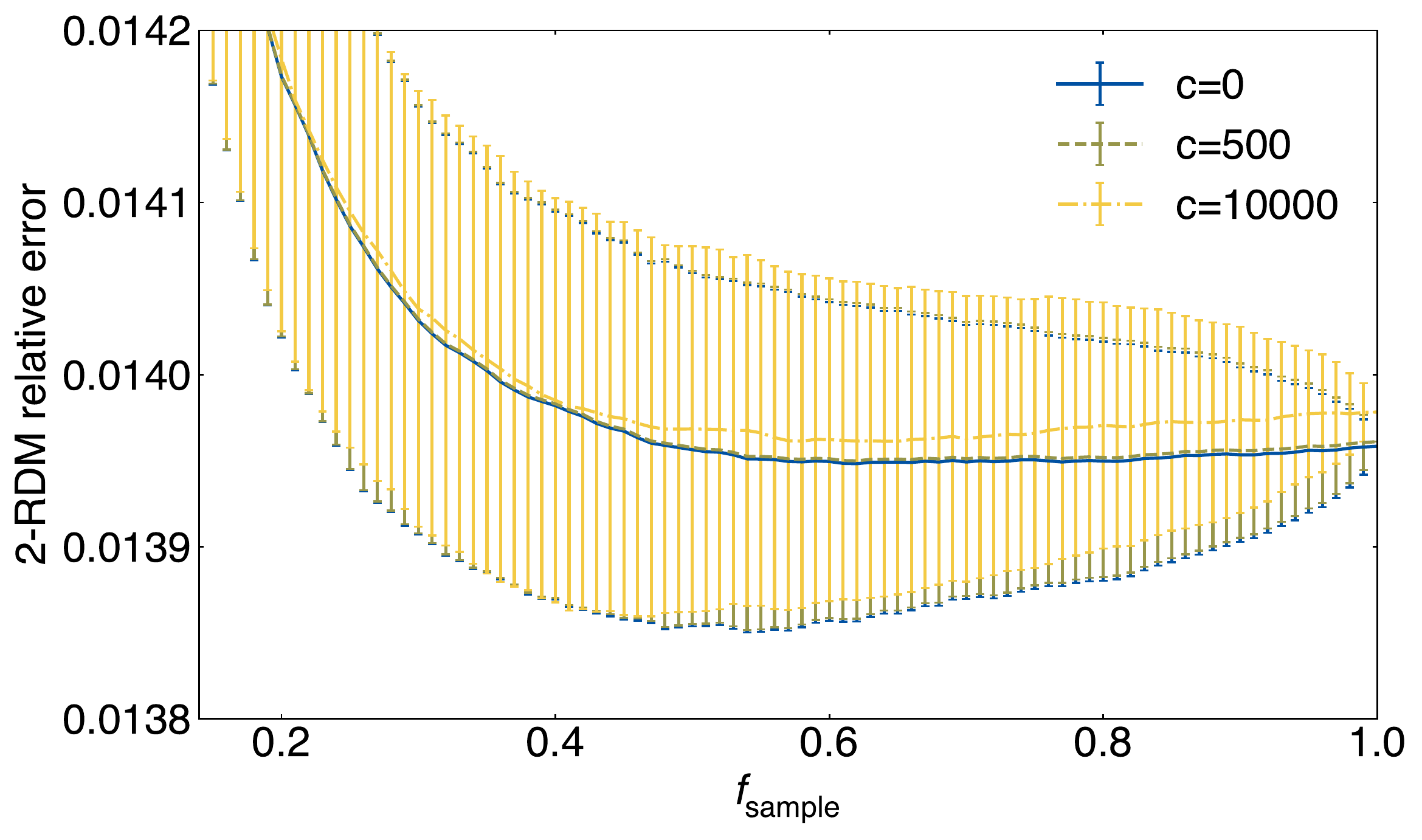}
    \caption{Completion errors of BeH CCSD $P^{\alpha\alpha\alpha\alpha}$ in the aug-cc-pVDZ basis in three settings, each with a different cost $c$ to switch the measurement, for a fixed number of shots per element (1317636). 
    The completion error at each $f_\text{sample}$ is averaged over errors from 1000 different random samplings and the standard deviation is taken as the error bar. }
    \label{figure:optimal}
\end{figure}

In Fig.~\ref{figure:optimal} we illustrate a typical result from searching for the optimal $N_\text{sample}$. We see that in this problem, for a given $m$ ($c=0$ line) it is in fact optimal to sample close to 100\% of the elements. This means matrix completion is performing almost entirely as a low-rank noise filter. In the other lines, we illustrate how the cost balance changes if we introduce a cost to switch the measurement setting when changing the Pauli strings (in multiples of the measurement cost, $c=500, 10000$ data; $c=500$ corresponds to the reported cost to switch measurements for the Sycamore quantum processor~\cite{sung2020using}). For a very high measurement setting cost, e.g. the case of $c=10000$, there is a benefit to sample fewer elements. However, in the subsequent calculations, we will neglect the cost of measurement switching and determine the best $m, f_\mathrm{sample}$ pair assuming $c=0$.

\subsection{Post-processing the completed 2-RDM}
\label{sec:post-process}

We can improve the results of matrix completion and noise filtering through post-processing. In the noisy quantum setting, this can be viewed as a form of error mitigation. We perform the following steps:
\begin{enumerate}
    \item For noiseless completion, we replace sampled terms in the completed 2-RDMs with their exact values  (i.e. giving zero completion error on the sampled terms). 
    \item We normalize $P$ (for the 3 spin components separately). 
    \item For noiseless completion, we apply matrix completion to obtain $\tilde{P}^r_M$, and the 2-RDM error of the model $P_M - \tilde{P}^r_M$ is added to our completed $\tilde{P}^r$. 
    \end{enumerate}

\section{Computational details}
\label{sec:computationaldetails}
We use 24 small molecular systems from the G2-1 test set~\cite{curtiss1991gaussian}, including both singlet (s) and triplet (t) states of \ce{CH2} and \ce{SiH2} and 20 other molecules in their lowest spin state
for our noiseless completion studies, and a further subset of the 7 smallest ones to study the basis dependence of noiseless completion and for the noisy completion studies. 
(The 7 smallest molecules serve as representative examples for the larger set, encompassing both those with the smallest and largest cost reductions in the noiseless completion results). We used (aug-)cc-pVXZ bases~\cite{dunning1989gaussian,woon1993gaussian,kendall1992electron} throughout and in CCSD and MP2 calculations froze the lowest energy orbitals (1$s$ for first row, 1$s$2$s$2$p$ for second row). Completion is thus only performed for the non-core part of $P$. Molecular geometries 
at the B3LYP/6-31G(2\textit{df, p})~\cite{becke1988density,lee1988development,becke1993density} level of theory 
were taken from Ref.~\cite{haunschild2012new, haunschild2009local}.  We computed unrestricted CCSD density matrices, as the reference ``exact'' density matrices. We  computed unrelaxed unrestricted 
 MP2 density matrices  as the model density matrices.

For our completion studies, we used a target completion error of $\epsilon_0 = 1\%$.  Minimizing the geometric coherence corresponds to minimizing
\begin{equation}
    \max_{p,q}\sum_s (C_{pi}C_{qk}U_{ik,s})^2
\end{equation}
where $U_{ik,s}$ are the singular vectors of $[P^r_M]_{ik,jl}$ and $C$ is the basis rotation matrix to be optimized. 
However, this minimization is numerically inconvenient because the $\max$ function is not differentiable everywhere. Instead, we perform the minimization
\begin{equation}
    \argmin_C \sum_{p,q}\left(\sum_s (C_{pi}C_{qk}U_{ik,s})^2\right)^4
\end{equation}
starting from 10 Haar random~\cite{haar1933massbegriff} initial guesses of orthogonal matrices $C$. In the noiseless matrix completion, for the MP2 $P_M$, we randomly generated 10 different element samplings for each $f_\text{sample}$, and for each sampling, used a maximum of 15000 iterations in the matrix completion optimization with the L-BFGS-B algorithm~\cite{liu1989limited,zhu1997algorithm}. $f_\text{sample}$ was chosen so that $\epsilon \leq \epsilon_0$ in no less than 90\% of the model completions. 
The CCSD 2-RDM completion was carried out using the same 10 element samplings as for MP2 2-RDMs. The CCSD completion errors were then averaged over all 10 trials except for cases where the optimization was not converged. In the noisy measurement setting, 10 random samplings were generated for each $f_\text{sample}$ to estimate the best $m, f_\mathrm{sample}$ pair.

All quantum chemistry calculations were carried out with PySCF~\cite{sun2018pyscf, sun2020recent}, while the conversion of the fermionic operators to Jordan-Wigner form was carried out using the OpenFermion package and the OpenFermion-PySCF plugin~\cite{mcclean2020openfermion, sun2018pyscf}. The Jordan-Wigner transformations were carried out on the 2-RDM in the coherence minimized basis.

\section{Results}

\subsection{Noiseless completion}

\label{sec:exact measurement results}

\begin{figure} 
    \centering
    \includegraphics[width=\columnwidth]{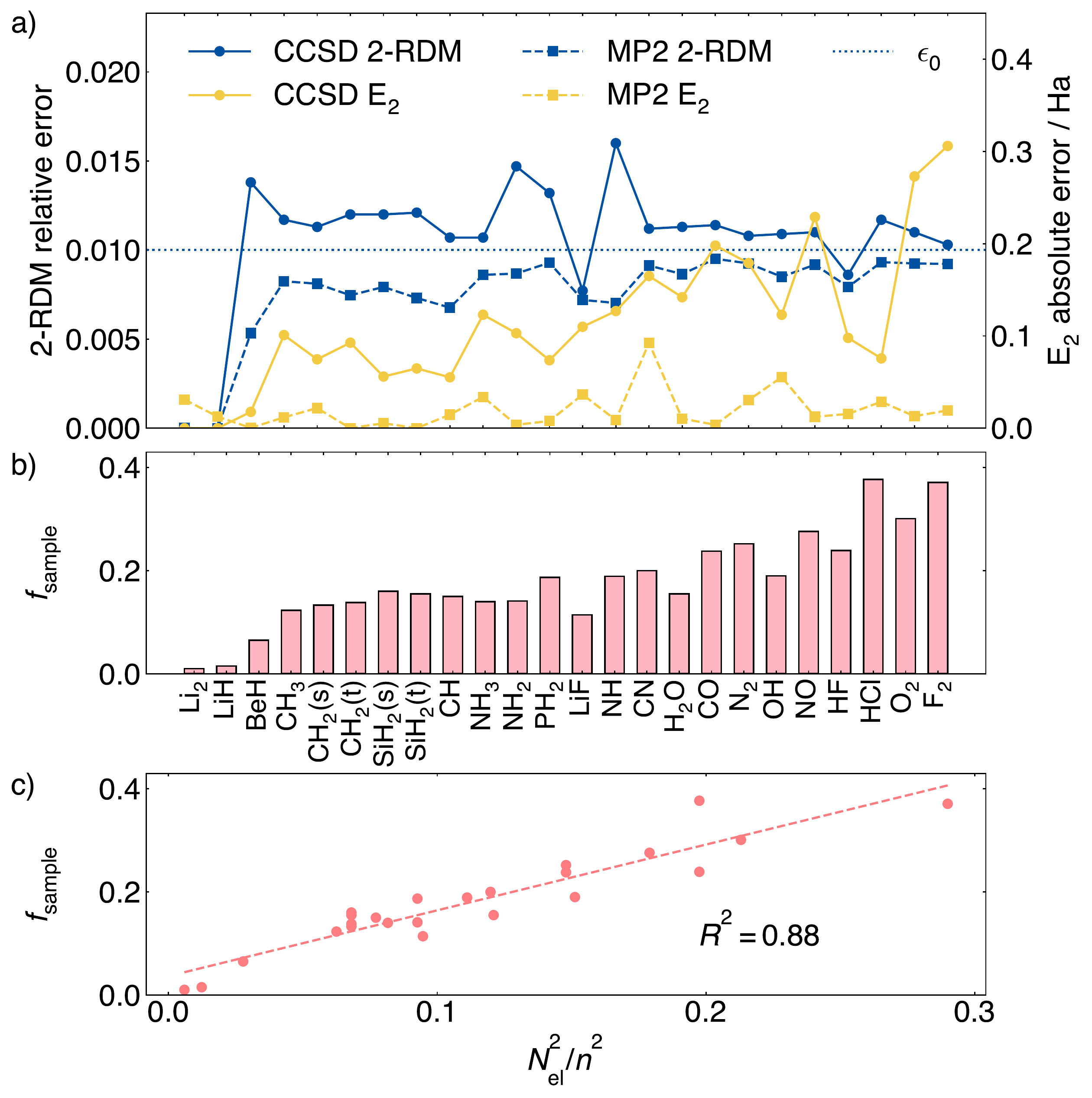}
    \caption{\textbf{a)} Using MP2 model $P_M$ to complete the CCSD 2-RDM. A target completion error $\epsilon_0=1\%$ for $P_M$  achieves  $\epsilon \approx 1\%$ in the completed CCSD 2-RDM. Also reported is the absolute error in the two-particle energy $E_2$.
    \textbf{b)} Fraction of fermionic terms sampled $f_\text{sample}$ used to complete 2-RDMs to the accuracy in \textbf{a)}.
     \textbf{c)} $f_\text{sample}$ is roughly proportional to $(N_\text{el}/n)^2$. }
    \label{figure:obsNeed}
\end{figure}
Fig.~\ref{figure:obsNeed}a shows the completion results for $\epsilon_0 = 1\%$ for 2-RDMs in the cc-pVDZ basis for the 24 systems, showing both the 2-RDM error and two-particle energy ($E_2$, for the non-core part of $P$) error for the target 2-PDMs from CCSD, as well as the 2-RDM and two-particle energy error for the MP2 model 2-PDMs. For the MP2 model quantities, $\epsilon< 1\%$, by design. (We note that LiH and Li$_2$ have anomalously small errors because their RDMs are approximately rank-1).
Across the series of molecules, this translates to approximately 0.02 Ha error in the MP2 two-particle energy.

The error of the MP2 model translates to the observed errors in the target CCSD 2-PDM matrix completion, with 
 $\epsilon \leq 1.5\%$. The CCSD two-particle energy error is somewhat larger and grows from left to right in the plot. The molecules in the plot are ordered in terms of increasing $N_\text{el}/n$. 
For molecules such as BeH, where there is a significant difference between the singular value of the MP2 model and CCSD (reflecting the stronger correlation described by CCSD), the main reason for the increased CCSD completion error comes from the increased rank truncation error.
In Fig.~\ref{figure:obsNeed}c, we see that $f_\text{sample}$ is roughly proportional to $(N_\text{el}/n)^2$. This comes from the rank of the Hartree-Fock 2-RDM as discussed in section~\ref{sec:2rdm}.

\begin{figure} 
    \centering
    \includegraphics[width=0.9\columnwidth]{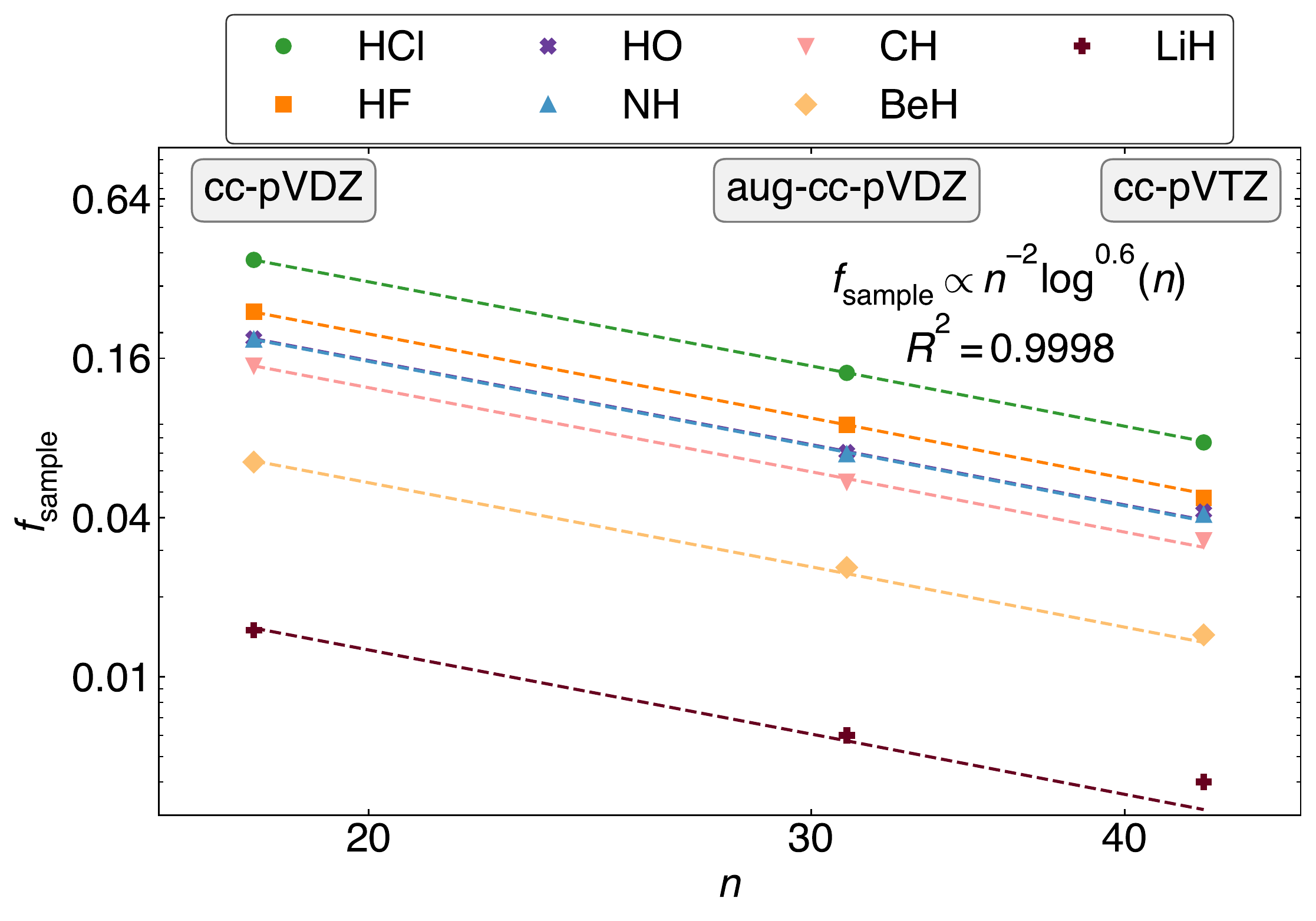}
    \caption{Fraction of elements sampled $f_\text{sample}$  ($\epsilon_0 = 1\%$) as a function of basis size for a set of 7 molecules.}
    \label{figure:basis}
\end{figure}

The above suggests that matrix completion is more useful in larger basis sets. We test this in the subset of 7 molecules in 
Fig.~\ref{figure:basis}. To achieve 1\% completion error in the model MP2 2-RDM, we find that the fraction of samples needed decreases with basis size as $\sim 1/n^2\log^{0.6}(n)$. This falls between the fundamental information lower bound of $1/n^2$~\cite{flammia2012quantum, haah2016sample}, below which $P$ is underdetermined, and the best provable bound of $1/n^2 \log(n)$, which guarantees a high probability to complete $P$ to a constant error~\cite{guctua2020fast}.

\begin{figure}[htbp]
    \centering
    \includegraphics[width=0.9\columnwidth]{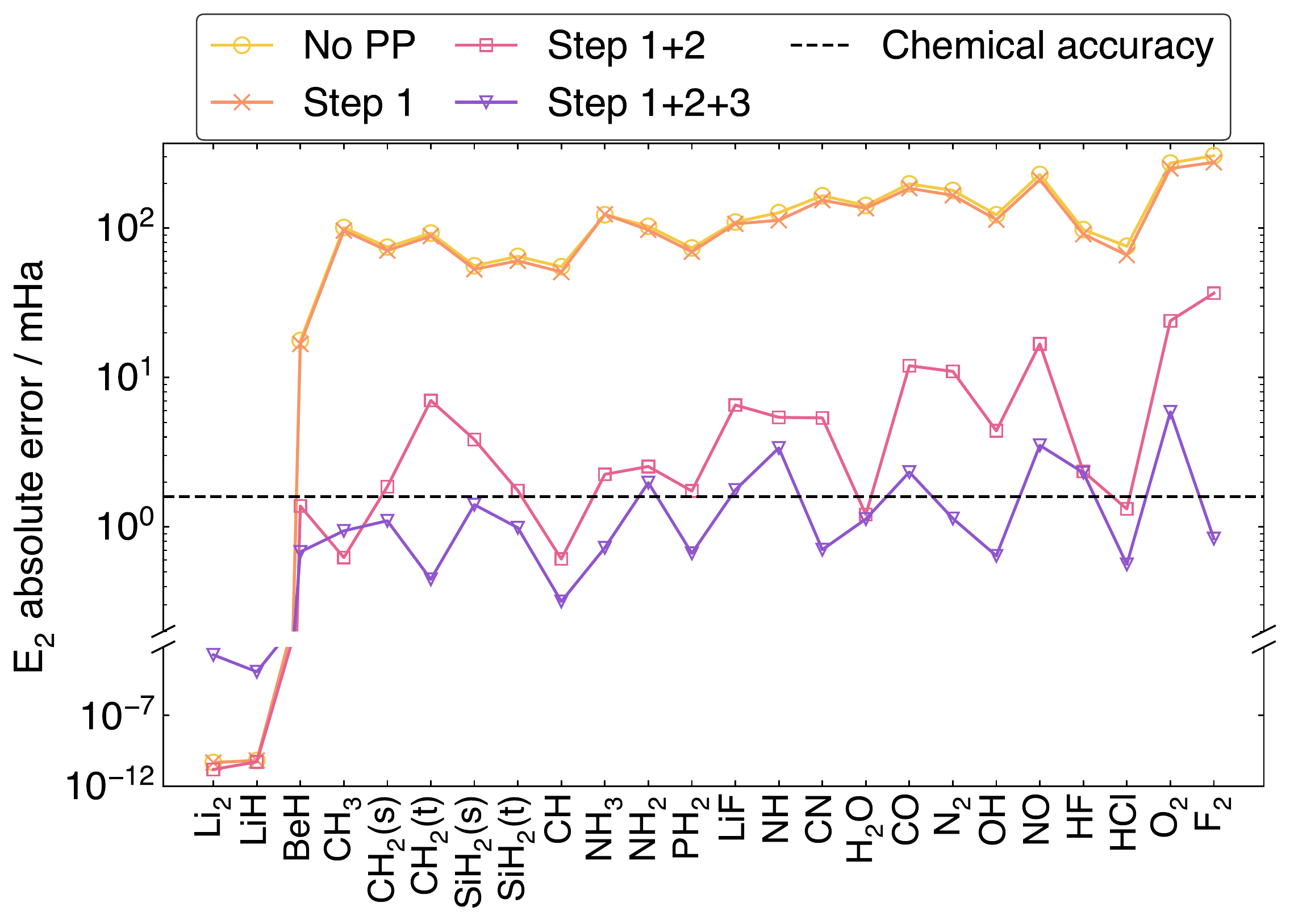}
    \caption{The effect on the two-particle energy $E_2$ of each post-processing option as defined in~Sec.~\ref{sec:post-process}.}
    \label{figure:PP}
\end{figure}

As expected, post-processing reduces the energy error of the completed target 2-RDMs. In Fig.~\ref{figure:PP} we show the effect of the different steps on the two-particle energy $E_2$. Out of the 3 post-processing steps, normalizing the trace of $P$ reduces the error the most, by 1-2 orders of magnitude. 
This suggests that the majority of energy error comes from the low-rank approximation, i.e. from  truncating small eigenvalues, which reduces the trace of $P$. After all post-processing steps, the two-particle energy errors are around chemical accuracy (1.6 mHa). 


\subsection{Noisy 2-RDM completion}
\label{sec:noisy measurement results}

We now carry out similar numerical experiments in the presence of measurement noise. 
In Fig.~\ref{figure:costReduction}a, we report the average number of shots (i.e. number of shots divided by the number of unique elements of $P$, denoted $\bar{m}$) in the standard measurement scheme and the average number of shots 
in the matrix completion scheme required to obtain $\epsilon_0 = 1\%$ for the subset of 7 molecules in the aug-cc-pVDZ basis.
We see that across all molecules, there is a significant reduction ($1/f_m$) in the average number of shots required compared to the standard measurement scheme; the total reduction is between 1 to 3 orders of magnitude in the aug-cc-pVDZ basis. 
For matrix completion, the associated $f_\mathrm{sample}$ 
used to generate the matrix completion data in Fig.~\ref{figure:costReduction}a is reported in Fig.~\ref{figure:costReduction}b. 
Almost all terms are measured for all the molecules, consistent with Fig.~\ref{figure:optimal}. Thus, resource reduction primarily comes from filtering the statistical noise in the measurements.
In Fig.~\ref{figure:costReduction}c we show the observed measurement cost reduction $1/f_m$ to complete to 1\% accuracy as a function of the estimated rank of $P_M$. We find $f_m \sim r/d$ (the relative rank of the 2-RDM), which, when rescaled for $||P|| \sim r$, matches theoretical sample complexities of low-rank completions performed on normalized density matrices $||\bar{P}||=1$~\cite{flammia2012quantum,guctua2020fast}. Therefore, resource reduction due to high-rank noise filtering is closely related to the low-rank property of the 2-RDM. 

\begin{figure}[htbp]
    \centering
    \includegraphics[width=0.9\columnwidth]{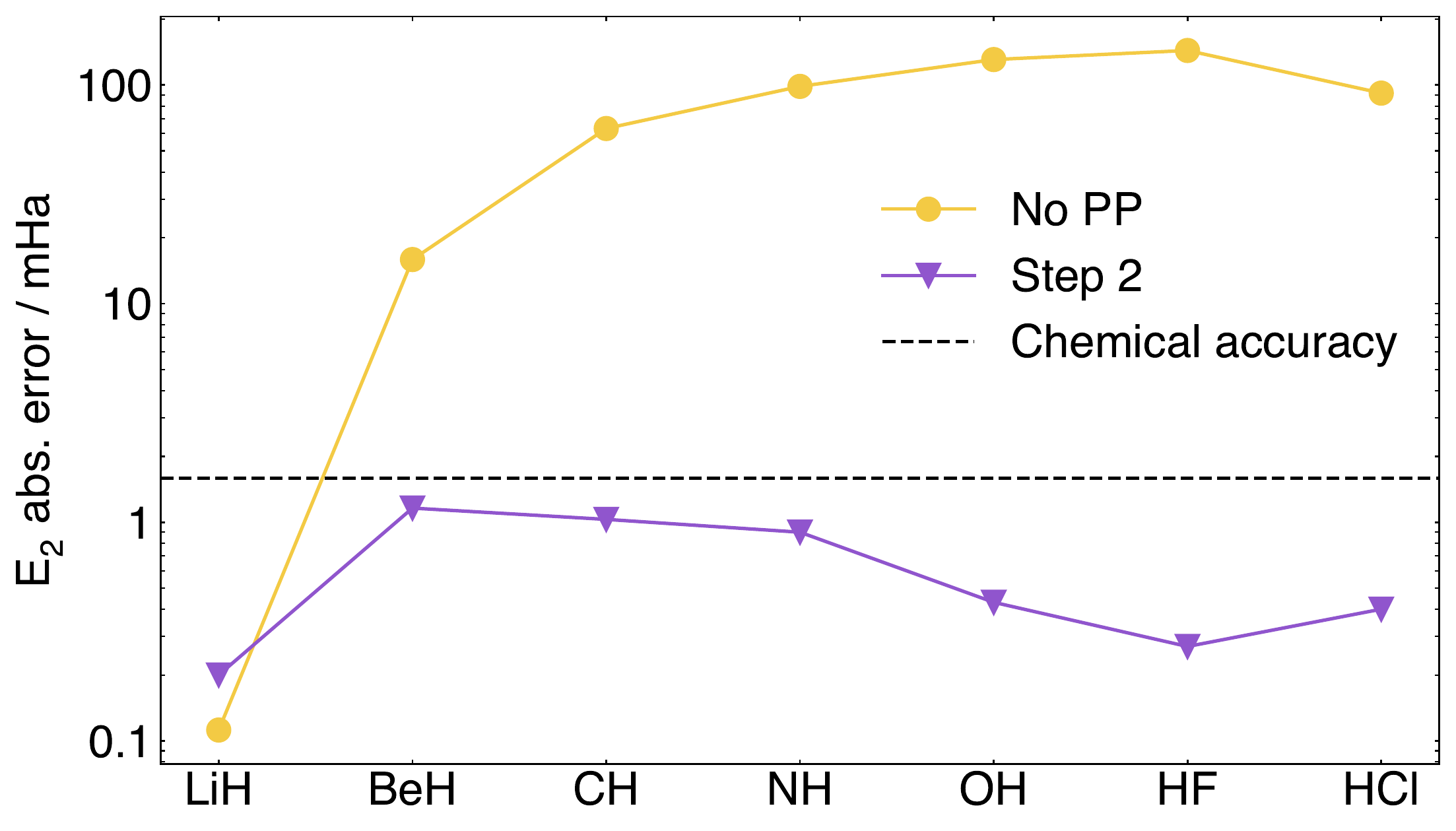}
    \caption{Absolute energy error from the completed 2-RDM before and after step~2 (trace normalization) of post-processing (PP). After post-processing, the two-particle energy error is within chemical accuracy (1.6 mHa).}
    \label{figure:noisyEnergy}
\end{figure}

In Fig.~\ref{figure:noisyEnergy}, we show the energy error and 2-RDM error before and after normalizing $P$ to the correct number of electrons. After this post-processing, the energy errors are all within chemical accuracy. 

\begin{figure}[htbp]
    \centering
    \includegraphics[width=\columnwidth]{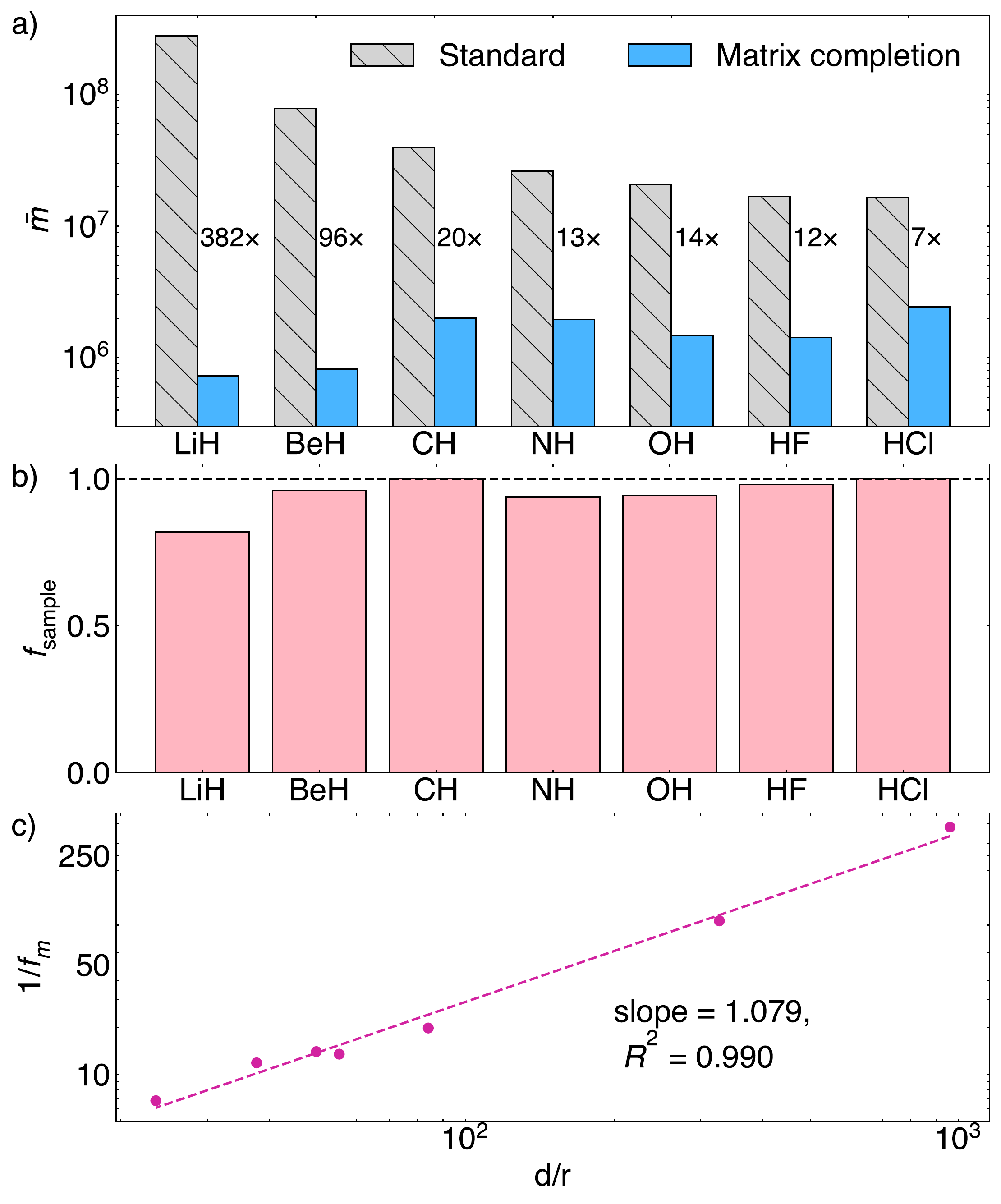}
    \label{figure:costReduction}
    \caption{\textbf{a)} 
    The average number of shots per unique Pauli term $\bar{m}$ for 7 molecules needed in the ``standard'' measurement scheme and when using ``matrix completion''. Their ratio $1/f_m$, i.e. the factor of measurement cost reduction, is reported next to the bars.   
    \textbf{b)} Fraction of fermionic terms sampled $f_\text{sample}$ used in \textbf{a)} ``matrix completion''. \textbf{c)} The  measurement cost reduction, $1/f_m$, is proportional to $d/r$, where $r$ is the MP2 rank estimate used in matrix completion. ($d$ and $r$ are averaged over the 3 spin sectors). }
    \label{figure:costReduction}
\end{figure}

\section{Conclusions}

We have demonstrated that matrix completions can effectively reduce the effort to obtain fermionic $k$-RDMs of interesting in electronic structure, and in particular, the 2-RDM. This was achieved by exploiting the low-rank structure as well as information obtained from approximate models of the 2-RDM. 

The current work has immediate applications in both classical and quantum algorithms to obtain 2-RDMs. In the classical setting, we envision that these techniques can easily be employed in quantum Monte Carlo simulations. In the hybrid quantum algorithm setting, there exist other techniques to reduce the measurement resources, such as optimizing the groups of qubit-wise commuting Pauli terms~\cite{verteletskyi2020measurement,bonet2020nearly}, or employing classical shadows~\cite{huang2020predicting, zhao2021fermionic}. It is likely these methods can be employed in conjunction with the matrix completion technique. In addition, it will be interesting to explore analogs of matrix completion which use the tensor structure of the 2-RDM, or impose additional constraints, such as $N$-representability conditions~\cite{coleman2000reduced}.

\section{Acknowledgement}
We thank Johnnie Gray, Yu Tong, Zhi-Hao Cui, Xuecheng Tao, Oscar F. Leong, Hsin-Yuan Huang, Steven T. Flammia, Shumao Zhang, Mario Motta, Shi-Ning Sun, Antonio Mezzacapo, and Scott E. Smart for helpful discussions. L. P. was supported by the  Center for Molecular Magnetic
Quantum Materials, an Energy Frontier Research Center funded
by the U.S. Department of Energy, Office of Science, Basic
Energy Sciences under award no. DE-SC0019330. X. Z. was supported by the U.S. Department of Energy, Office of Science, under award no. DE-SC0019374.
\textbf{Author contributions:} L. P. and G.K.-L. C. designed the study and wrote the manuscript. L. P. performed the calculations. L. P., X. Z., and G.K.-L. C. contributed to the writing and editing of the manuscript. 

\clearpage
\bibliographystyle{apsrev4-2}
\bibliography{reference}
\end{document}